\documentclass[aip,jcp,reprint,amsmath,amssymb,floatfix,citeautoscript]{revtex4-1}
\usepackage{amsmath}
\usepackage[usenames,dvipsnames]{color}
\usepackage{amssymb}
\usepackage{braket}
\usepackage{graphicx}
\DeclareMathOperator{\Tr}{Tr}

\usepackage{ragged2e}
\usepackage[labelsep=period,format=plain,justification=justified,font=footnotesize]{caption}
\DeclareCaptionJustification{myjust}{\justifying}
\usepackage{times}
\usepackage{txfonts}

\begin{document}

\title{Extending the applicability of Redfield theories into highly non-Markovian regimes}
\author{Andr\'{e}s Montoya--Castillo}
\affiliation{Department of Chemistry, Columbia University, New York, New York, 10027, USA}
\author{Timothy C. Berkelbach}
\affiliation{Princeton Center for Theoretical Science, Princeton University, Princeton, New Jersey, 08544, USA}
\author{David R. Reichman}
\affiliation{Department of Chemistry, Columbia University, New York, New York, 10027, USA}

\date{\today}

\begin{abstract}
We present a new, computationally inexpensive method for the calculation of
reduced density matrix dynamics for systems with a potentially large number of
subsystem degrees of freedom coupled to a generic bath.  The approach consists
of propagation of weak--coupling Redfield--like equations for the high frequency
bath degrees of freedom only, while the low frequency bath modes are
dynamically arrested but statistically sampled.  We examine the improvements
afforded by this approximation by comparing with exact results for the
spin--boson model over a wide range of parameter space. The results from the
method are found to dramatically improve Redfield dynamics in highly
non--Markovian regimes, at a similar computational cost.  Relaxation of the the
mode--freezing approximation via classical (Ehrenfest) evolution of the low
frequency modes results in a dynamical hybrid method.  We find that this
Redfield--based dynamical hybrid approach, which is computationally more
expensive than bare Redfield dynamics, yields only a marginal improvement over
the simpler approximation of complete mode arrest.  
\end{abstract}

\maketitle

\normalsize

\section{Introduction}
\label{Sec:Introduction}
 
Useful approximate methods for the description of quantum dynamics and
relaxation can often circumvent the large computational expense of numerically
exact approaches while maintaining quantitative accuracy in certain regions of
parameter space.  The general applicability of such methods, however, is often
limited and their domain of validity difficult to assess.  The most widely used
approximate approaches fall into two broad and general classes.  The first
class of methods involves techniques that employ uncontrolled approximations to
yield dynamics which are non--perturbative in the various couplings (e.g.,
intra--system or system--bath) that characterize the problem. The second class
of methods are systematically perturbative in a well--defined coupling
parameter, but are free from further classical or semiclassical approximations,
at least for simply defined models such as a the spin--boson model.   

One of the most celebrated perturbative techniques is a lowest--order treatment
of the system--bath coupling, known traditionally as Redfield
theory.\cite{Bloch1957,Redfield1965,Nitzan}  As we will show later, the relevant
dimensionless parameter characterizing the accuracy of Redfield theory is $\eta
= \mathrm{max}\left[2\lambda/\beta \omega_c^2, 2\lambda/\pi \omega_c\right]$, 
where $\lambda$ is the nuclear reorganization energy, $\beta =
1/k_BT$ is the inverse temperature, and $\omega_c$ is a characteristic bath
frequency (we henceforth work in units with $\hbar = 1$).  Redfield theory
becomes unreliable when $\eta \gtrsim 1$.  We emphasize that $\eta$ is
controlled by multiple bath parameters and, in particular, low frequency
degrees of freedom (small $\omega_c$) limit the range of accessible
reorganization energies.  Indeed, violations of this condition explain the
failures of Redfield theory found by Ishizaki and Fleming\cite{Ishizaki2009b}
for certain models of excitation energy transfer, which appear to be
characterized by low--frequency protein baths.

While the very lowest frequency degrees of freedom are thus most problematic
for Redfield theory to handle (even in its non--Markovian forms), it is often
the case that nuclear modes of such frequencies are effectively frozen on the
time scale of relevance for the system's dynamics.  In this regard, the key
function of such modes is simply to provide static energetic disorder for the
more rapidly evolving degrees of freedom.  This suggests a methodology whereby
the very low frequency phonons are approximated as static (and treated
\textit{non--perturbatively} as a source of static disorder), while the
remaining portion of the bath is treated dynamically within Redfield theory.
Here we develop this ``Redfield theory with frozen modes" (Redfield--FM)
method, and show that it greatly extends the applicability of Redfield theory
into highly non--Markovian dynamical regimes at essentially no change in
computational cost.

The outline of this paper is as follows.  In Sec.~\ref{Sec:Theory}, we
introduce the theoretical background for the Redfield equations and the
derivation and general properties of the Redfield--FM extension.  In addition,
we also introduce the spin--boson Hamiltonian as the model system on which we
test the methods developed in this paper.
Section~\ref{Sec:Results:Computational} presents the computational details in
the implementation of the Redfield--FM method, while
Sec.~\ref{Sec:Results:RedfieldFM} presents illustrative results for the method.
In Sec.~\ref{Sec:Results:Hybrid}, we relax the mode--freezing approximation via
the derivation and implementation of a dynamical hybrid method
(hybrid--Redfield) that combines Redfield dynamics for the high--frequency part
of the bath coupled to the electronic system and Ehrenfest dynamics for the
low--frequency modes.  In Sec.~\ref{Sec:Conclusions}, we conclude.  

\section{Theory}
\label{Sec:Theory}

\subsection{Model} 
\label{Sec:Theory:Model}

First, we briefly describe the model system we use to test the approximations
developed in subsequent sections.  This allows us to define notation and
parameters that will be used in our numerical comparisons.   We focus on the
well--known spin--boson model, which consists of a two--level system coupled
linearly to a harmonic bath.  This model has been extensively used to
investigate a wide variety of relaxation, charge and energy transport processes
in condensed phase systems.\cite{WeissQDS}  

The total Hamiltonian is divided into system, bath, and interaction components, $H =
H_{\mathrm{sys}} + H_{\mathrm{bath}} + V$. The system Hamiltonian takes the form 
    \begin{equation} 
    H_{\mathrm{sys}} = \varepsilon\sigma_z + \Delta \sigma_x, 
    \end{equation} 
where $\sigma_i$, $i = \{ x, y, z\}$, are the Pauli matrices, $2\varepsilon$ is
the energy difference, and $\Delta$ is the coupling between the two
electronic sites, which is here assumed to be static.  The bath Hamiltonian
consists of an infinite set of harmonic oscillators,
    \begin{equation} 
    H_{\mathrm{bath}} = \sum_k \frac{1}{2}\left[P_k^2 + \omega_{k}^2Q_{k}^2\right].  
    \end{equation} 
Lastly, the system--bath interaction couples the electronic states linearly to
coordinates of the bath oscillators, 
    \begin{equation} \label{Eq:SBCoupling}
    V = \sigma_z\sum_{k}c_{k}Q_{k}.  
    \end{equation}    
Physically, the system--bath coupling acts as a (quantum) fluctuating field
that shifts the origin of the bath harmonic oscillators by a magnitude that
depends on the system's electronic state and the strength of the coupling.   

The spectral density, which completely determines the coupling between the bath
and the system, is taken to be Ohmic with a Lorentzian cutoff (Debye form),  
    \begin{equation}\label{eq:DebyeSD}
    J(\omega) = \frac{\pi}{2}\sum_k \frac{c_{k}^2}{\omega_{k}}\delta(\omega - \omega_{k})
    = 2\lambda \omega_c \frac{\omega}{\omega^2 + \omega_c^2}. 
    \end{equation}
The cutoff frequency, $\omega_c$, characterizes how quickly the bath relaxes
toward equilibrium, while the reorganization energy, $\lambda =
\pi^{-1}\int_0^{\infty} d\omega \  J(\omega)/\omega$, characterizes the energy
dissipated by the environment after a Franck--Condon transition between
electronic states.  It is important to note that the methods studied
here are neither limited to the spin--boson model nor to the Debye form for the
spectral density.

\subsection{Time--local Redfield Dynamics}
\label{Sec:Theory:Redfield}

Because of its simplicity in the time domain we employ the time--local (i.e.
time--convolutionless) form of the generalized Redfield equations.  A full
derivation of these equations is contained in Appendix \ref{App:Redfield}.  Here our
aim is to highlight important but often overlooked aspects pertaining to the
applicability of the Redfield approach.  For the spin--boson model, the
time--local version of the Redfield theory takes the following form,
    \begin{equation}\label{Eq:SchroDensityOperatorRedfield}
    \begin{split}
    \frac{d}{dt} \rho(t) &= -i \left[H_{\mathrm{sys}}, \rho(t)\right]  \\
    &\hspace{1em} - \int_0^{t} d\tau\ \Big\{ C(\tau)
        \left[\sigma_z(0), \sigma_z(-\tau)\rho(t)\right] \\
    &\hspace{5em} - C^*(\tau)\left[\sigma_z(0), \rho(t)\sigma_z(-\tau)\right]\Big\},
    \end{split}
    \end{equation}
where all operators except the reduced density matrix (RDM) are evolved in the 
interaction picture,
$O(t) = e^{-i(H_{\mathrm{sys}} + H_{\mathrm{bath}})t}Oe^{i(H_{\mathrm{sys}} +
H_{\mathrm{bath}})t}$, and the free bath correlation function is given by
    \begin{equation} \label{Eq:BathCorrelationFunction}
    \begin{split}
    C(t) &= \sum_{k}c_k^2 \Tr_{\mathrm{bath}} \Big\{ \rho_{\mathrm{bath}} Q_k(t)Q_k(0) \Big\}\\
    	&= \frac{1}{\pi} \int_0^{\infty} d\omega  J(\omega)[\coth(\beta\omega/2)\cos(\omega t) -
        i\sin(\omega t)].
    	\end{split}
    \end{equation}  
By going to the interaction picture with respect to
$H_{\mathrm{sys}}+H_{\mathrm{bath}}$ (to eliminate the free-evolution) and
formally integrating the equation of motion
(\ref{Eq:SchroDensityOperatorRedfield}), one finds
    \begin{equation}\label{eq:InteractionPicExp}
    \rho_I(t) = \rho_I(0) \left[1 + O\left(\int_0^t d\tau_1 \int_0^{\tau_1} d\tau_2\ C(\tau_2) \right) + \dots \right].
    \end{equation}
    
Examination of the function $\int_0^td\tau_1 \int_0^{\tau_1}d\tau_2\ C(\tau_2)$
reveals the natural dimensionless parameter that determines the limit of
validity of Redfield theory.  In general, even in the non--Markovian case, we
expect that the ordering of terms in the expansion (\ref{eq:InteractionPicExp})
is governed by the function $\int_0^td\tau_1 \int_0^{\tau_1}d\tau_2\ C(\tau_2)
= \eta g(t)$, where $\eta$ is a dimensionless constant and $g(t)$ is a
function expressed in terms of a scaled, dimensionless time variable.  In the
high--temperature limit ($\beta \omega_c \ll 1$), where $C(t) \approx
(2\lambda/\beta) e^{-\omega_c t}$, it is easy to show
\begin{subequations}
\begin{align}
\eta &= 2\lambda/(\beta \omega_c^2) \\
g(t) &= e^{-\omega_c t} - 1 + \omega_c t.
\end{align}
\end{subequations}
In the low--temperature limit ($\beta \omega_c \gg 1$), we assume
that the low--frequency behavior of the spectral function dominates, so we
choose $J(\omega) = (2\lambda \omega /\omega_c)e^{-2\omega/\pi \omega_c}$ as
an approximation to the Debye form in Eq.~(\ref{eq:DebyeSD}) that exactly
matches the value of $\lambda$ and its low--frequency asymptotic behavior.
Using this spectral density, one can show 
\begin{subequations}
\begin{align}
\eta &= \frac{2\lambda}{\pi \omega_c} \\
g(t) &= \ln\left[ 1 + (\omega_c t)^2 \right]
    + 2 \ln\left[\frac{\sinh (\pi t/\beta \hbar)}{(\pi t/\beta\hbar)} \right].
\end{align}
\end{subequations}
Thus, for a Debye spectral density, Redfield theory will be reliable
as long as 
$\mathrm{max}\left[2\lambda/\beta \omega_c^2, 2\lambda/\pi \omega_c\right]$
is not significantly larger than unity.\footnote{This criterion is only approximately valid because it ignores the magnitude of the system timescales.  A more rigorous expression could be obtained via direct examination of the ratio of the 2$^{nd}$ and 4$^{th}$ order terms that may be found, e.g., in the work of Laird and Skinner\cite{Laird1991} or Ref.~\onlinecite{BreuerPetruccione}.}  It should be
noted that recent work purported to be in the Redfield limit actually violates
the above condition.\cite{Landry2015}
As long as the relevant energy scales in the system Hamiltonian are not
too large, we expect the above to hold.  In cases where the system's bare
energy difference $\varepsilon$ is the largest energy scale in the problem, the
dynamics will be mediated by multi--phonon processes which are a challenge for
lowest--order Redfield--like theories.  However, in this limit, the problem
acquires an increasing amount of `pure--dephasing' character, for which the
time--local version of Redfield theory provides an exact multi--phonon
resummation.

\subsection{Redfield Theory with Frozen Modes}
\label{Sec:Theory:RedfieldFM}

As discussed in the Introduction, low--frequency bath modes $\omega_k$ which
lead to a violation of the validity of the Redfield theory frequently evolve so
slowly as to be effectively static on the electronic timescale.  Here we
develop the ``Redfield theory with frozen modes" (Redfield--FM) method, based on
the physically appealing notion of dividing modes into a low--frequency portion
(treated as static disorder), and a high--frequency bath (treated by time--local
Redfield theory).  The separation of modes used here mirrors that utilized in
previous work on a F\"orster--like dynamical hybrid
approach.\cite{Berkelbach2012a}  The approach presented here is not in any
way limited to time--local Redfield theory (nor even to any specific flavor
of Redfield theory).  However, due to pathologies associated with a
strictly Markovian Redfield theory, we suggest certain adjustments to the
partitioning algorithm, as discussed in Appendix \ref{App:Markovian}.

First, it is advantageous to assume that the total density matrix is
multiplicatively separable into weakly interacting parts, i.e.,
$\rho_{\mathrm{tot}}(t) \approx
\rho_{\mathrm{slow}}(0)\rho_{\mathrm{sys+fast}}(t)$ where
$\rho_{\mathrm{slow}}(0)$ is the density matrix of the frozen low--frequency
``slow modes'' and $\rho_{\mathrm{sys+fast}}(t)$ is the density matrix for the
system and high--frequency ``fast modes''.  As in
Ref.~\onlinecite{Berkelbach2012a}, a splitting function, $S(\omega)$, divides
the spectral density into two components, $J(\omega) =
J_{\mathrm{slow}}(\omega) + J_{\mathrm{fast}}(\omega)$, where 
    \begin{equation} \label{Eq:Jslow}
    J_{\mathrm{slow}}(\omega) = S(\omega, \omega^*)J(\omega),
    \end{equation}
and
    \begin{equation}\label{Eq:Jfast}
    J_{\mathrm{fast}}(\omega) = [1-S(\omega, \omega^*)]J(\omega)
    \end{equation}
Here we take the same form of the splitting function as that suggested in
Ref.~\onlinecite{Berkelbach2012a}, namely,
    \begin{equation}
    S(\omega, \omega^*) = \left\{
      \begin{array}{lr}
       [1 - (\omega/\omega^*)^2]^2 \quad &: \quad \omega < \omega^*\ \\
       0  &: \quad \omega \geq \omega^*,
      \end{array}
    \right.
    \end{equation}
which, by virtue of its smoothness, avoids problems associated with long--time
oscillatory tails in the bath correlation function.\cite{Berkelbach2012}  While
the above splitting induces no errors if the dynamics are treated exactly, it
is clear that $\omega^*$ serves as a free parameter that allows one to tune the
optimum percentage of frozen bath modes, and hence the accuracy of the results
if the dynamics are treated within our approximate method.  The utility of the
present method is greatly enhanced if a physical \textit{a priori} prescription
for choosing $\omega^*$ based only on the parameters of the initial Hamiltonian
can be put forth.  In this work we choose $\omega^* = \omega_R/4$, 
where $\omega_R = 2 \sqrt{\varepsilon^2 + \Delta^2}$ is the
system Rabi frequency.  This choice for $\omega^*$ is 
simple, yields non--trivial improvements over standard
Redfield theory, and may easily be generalized to multiple electronic states.
Physically, this choice partitions the bath into modes that evolve
slower than the system (to be treated as frozen) and modes that evolve faster 
than the system (to be treated via Redfield theory). 
However, it should be noted that this choice is not always optimal.  Future
work will be devoted to the goal of arriving at an optimal choice of
$\omega^*$.  Other choices for $\omega^*$ that fit within the general physical
guidelines discussed above will be discussed in the Results section.  

With a prescription for choosing $\omega^*$ in hand, it is possible to separate
the fast and slow portions of the Hamiltonian, starting with the interaction,
$V_{\mathrm{fast}} = \sigma_z\sum_{k \in \mathrm{fast}}c_{k}Q_{k}$ and
$V_{\mathrm{slow}} = \sigma_z\sum_{k \in \mathrm{slow}}c_{k}Q_{k}$.  Regrouping
terms, it is evident that freezing the slow part, $V_{\mathrm{slow}}$, will
yield a classical reorganization energy that renormalizes the bias for every
realization of the bath's initial conditions. The modified total Hamiltonian
now takes the following form, 
    \begin{equation} \label{Eq:Hsc}
    \begin{split}
    H' &= [\varepsilon + \lambda^{cl}(0)]\sigma_z + \Delta\sigma_x\\
    &+ \sigma_z \sum_{k\in \mathrm{fast}}c_{k}Q_{k} +\frac{1}{2}\sum_{k}\Big[P_{k}^2 + \omega_{k}^2Q_{k}^2 \Big],
    \end{split}
    \end{equation}
where the classical reorganization energy is defined as $\lambda^{cl}(0)
=\sum_{k\in \mathrm{slow}} c_{k}Q_{k}(0)$ and the set of $Q_{k}(0)$ is sampled
from a bath distribution function after the discretization of $J_{\mathrm{slow}}(\omega)$.
Physically, each realization of the frozen bath degrees of freedom constitutes a local, rigid
environment that modifies the site energies for the system Hamiltonian.
The time--local Redfield dynamics, under the
Hamiltonian in Eq.~(\ref{Eq:Hsc}) with an interaction given by
$V_{\mathrm{fast}} = \sigma_z \sum_{k\in \mathrm{fast}}c_{k}Q_{k}$, are
subsequently ensemble averaged with respect to the slow frozen modes.  Thus,
there are two important differences for the Redfield equations used in each
realization of the frozen modes: (i) the bias is given by 
$\tilde{\varepsilon} \equiv \varepsilon + \lambda^{cl}(0)$, and (ii) the bath
correlation function given by Eq~(\ref{Eq:BathCorrelationFunction}) is
modified, with $J(\omega)$ replaced by $J_{\mathrm{fast}}(\omega)$.

To account for the classical frozen modes, the nonequilibrium population
dynamics takes the following form,
    \begin{equation}\label{Eq:PopDyFrozenModes}
    \langle \sigma_z(t)\rangle \approx \int d\mathbf{P}d\mathbf{Q}\ \rho_{\mathrm{slow}}(\mathbf{P}, \mathbf{Q}, 0) 
    \Tr_{\mathrm{sys+fast}}[\rho_{\mathrm{sys+fast}}(t)\sigma_z],
    \end{equation}
where $\rho_{\mathrm{slow}}(\mathbf{P}, \mathbf{Q}, 0)$ could be either the
classical distribution function or the Wigner transform
of the equilibrium density operator of the slow bath degrees of freedom, and
$\rho_{\mathrm{sys+fast}}(t)$ is the reduced density matrix of the system and
the fast bath degrees of freedom.  Observables such as 
$\Tr_{\mathrm{sys+fast}}[\rho_{\mathrm{sys+fast}}(t)\sigma_z]$ may then be
calculated via the Redfield equations.

\begin{figure}
\includegraphics[width=8.5cm]{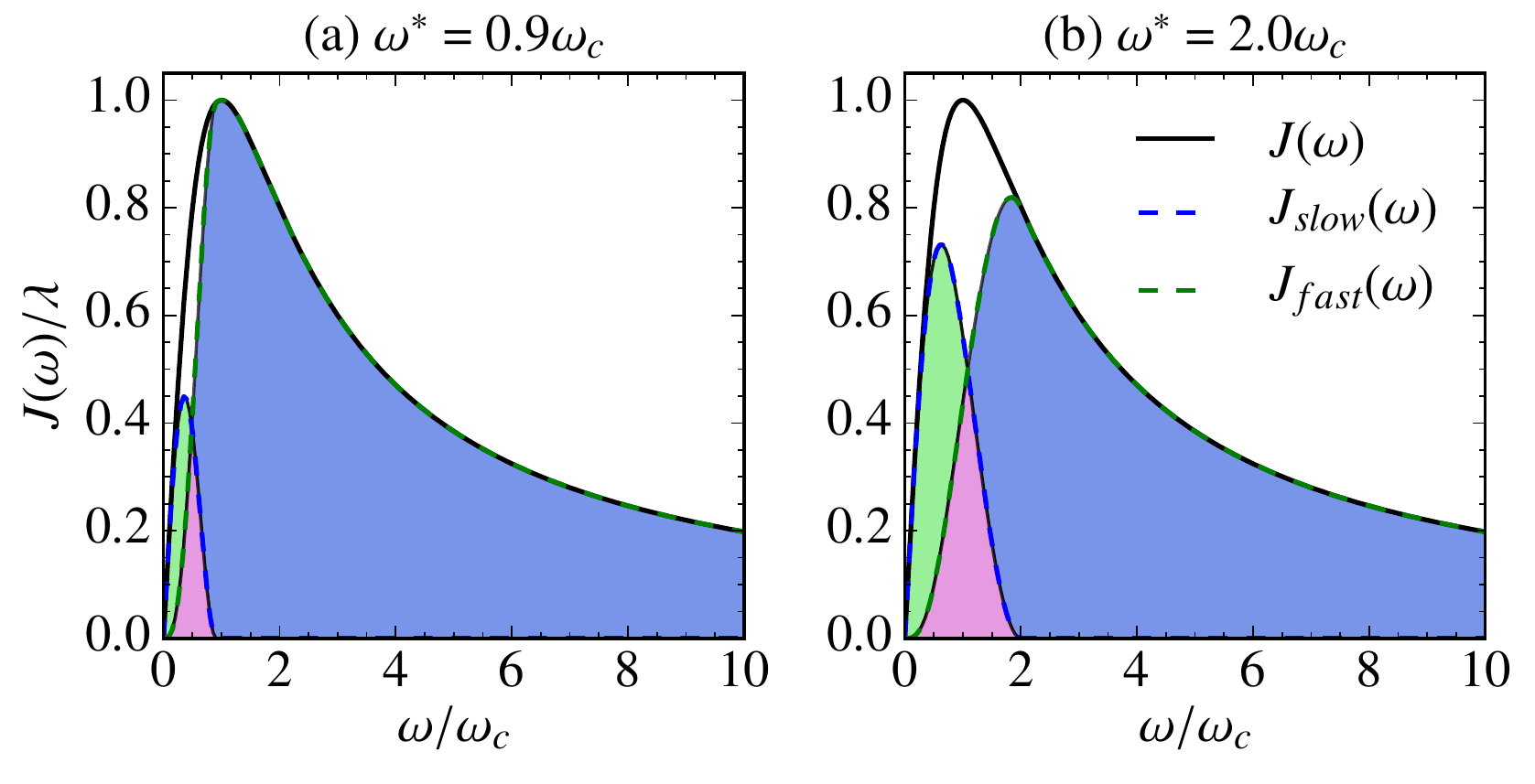} 
\caption{Spectral density and splitting via $S(\omega, \omega^*)$,
illustrating the two situations expressed in Eqs.~(\ref{Eq:Jslow}) and
(\ref{Eq:Jfast}).}
\end{figure}

To understand the relaxation processes with mode freezing for finite
$\omega^*$, it is first useful to investigate the effect of the approximation
at its most extreme, namely the adiabatic limit, where all bath modes are
assumed to be static ($\omega^* \rightarrow \infty$).  In this limit, we are
effectively in the Born--Oppenheimer regime, where excitations in the
electronic subspace move along the potential energy surface determined by the
frozen reservoir.  Analytical evaluation of the trace within
Eq.~(\ref{Eq:PopDyFrozenModes}) leads to the following expression for the
nonequilibrium population dynamics,
    \begin{equation}\label{Eq:PopDyn_allFrozenModes}
    \langle \sigma_z(t)\rangle \approx \int d\mathbf{P} d\mathbf{Q}\ \rho_{\mathrm{bath}}(\mathbf{P}, \mathbf{Q}, 0) \ \frac{\tilde{\varepsilon}^2  + \Delta^2\cos(\xi t)}{\tilde{\varepsilon}^2  + \Delta^2},
    \end{equation}
where $\xi = 2\sqrt{\tilde{\varepsilon}^2 + \Delta^2}$ is the Rabi frequency of the
modified system Hamiltonian.  The integration over the ensemble of equilibrium
configurations of the bath reduces to averaging over different values of
$\lambda^{cl}(0)$ that are consistent with the bath distribution function.  For
some realizations of the bath, Eq.~(\ref{Eq:PopDyn_allFrozenModes}) recovers the
Rabi oscillations characteristic of the isolated system if $\lambda^{cl}(0) =
0$.  Conversely, when $\lambda^{cl}(0) \neq 0$, the population starts from 1 at
$t = 0$ and will oscillate between $(\tilde{\varepsilon}^2  +
\Delta^2)/(\tilde{\varepsilon}^2  + \Delta^2) = 1$  and $(\tilde{\varepsilon}^2
- \Delta^2)/(\tilde{\varepsilon}^2  + \Delta^2)$.  Taking for simplicity
$\varepsilon = 0$, one notes that the lower bound of the population
oscillations increases with increasing $\lambda^{cl}(0)$, approaching 1 as
$\lambda^{cl}(0) \rightarrow \infty$.  This limit corresponds to an infinitely
rigid bath that completely localizes the excitation on its initial site.  

Averaging over different realizations of the slow modes decreases the amplitude
of oscillations in the population dynamics due to the decoherence between the
functions with distinct oscillation frequencies.  In general, Redfield theory
has difficulty describing non--Markovian, multi--step relaxation dynamics.
However, when $\lambda_{\mathrm{slow}} = \pi^{-1}\int_0^{\infty}d\omega\
J_{\mathrm{slow}}(\omega)/\omega$ is sufficiently large, the full dynamics produced by
Redfield theory with frozen modes at finite $\omega^*$ will include both a
slow, perhaps oscillatory component as well as a more rapid decay induced by
the high frequency modes in $J_{\mathrm{fast}}(\omega)$. These qualitative
considerations suggest this approach may correct certain deficiencies of conventional 
Redfield--like approaches.  In the next section, we test the approach quantitatively. 

\section{Results}
\label{Sec:Results}

In the following, we compare the numerically exact population dynamics reported
by Thoss \textit{et al.}\cite{Thoss2001} for the spin--boson model with a Debye
spectral density and the initial condition $\Gamma(0) =
\ket{1}\bra{1}\exp(-\beta H_{\mathrm{bath}})/Z_{\mathrm{bath}}$ with the
results obtained from the Redfield--FM method.      Subsequently, we examine
the effect of relaxing the mode--freezing approximation by treating the
low--frequency modes dynamically via the Ehrenfest method.  We call this latter
approach the hybrid--Redfield method, in analogy with the previously developed
hybrid--NIBA method.\cite{Berkelbach2012, Berkelbach2012a}  

\subsection{Computational Details}
\label{Sec:Results:Computational}

To treat the frozen portion of the spectral density, $J_{\mathrm{slow}}(\omega)$, we
have discretized the bath into $f = 300$ modes with frequencies and couplings
given by\cite{Craig2005, Berkelbach2012}
    \begin{equation}
    \omega_k = \omega_c \tan\Bigg[ \frac{\pi}{2f}(k - 1/2) \Bigg],
    \end{equation}	 
and
    \begin{equation}
    c_k^2 = \frac{2\lambda}{f}\omega_k^2.  
    \end{equation}

Initial conditions for the reservoir of frozen modes were sampled from a Wigner
distribution.  Sampling from this distribution becomes particularly important
at very low temperatures, where quantum effects become significant.  However, for
most cases, sampling from a Boltzmann distribution is sufficient since the
modes being samples are always low--frequency.  For
convergence, up to $10^4$ trajectories have been run for the results presented.  

\subsection{Redfield--FM Method}
\label{Sec:Results:RedfieldFM}

\begin{figure}
\includegraphics[width=8.5cm]{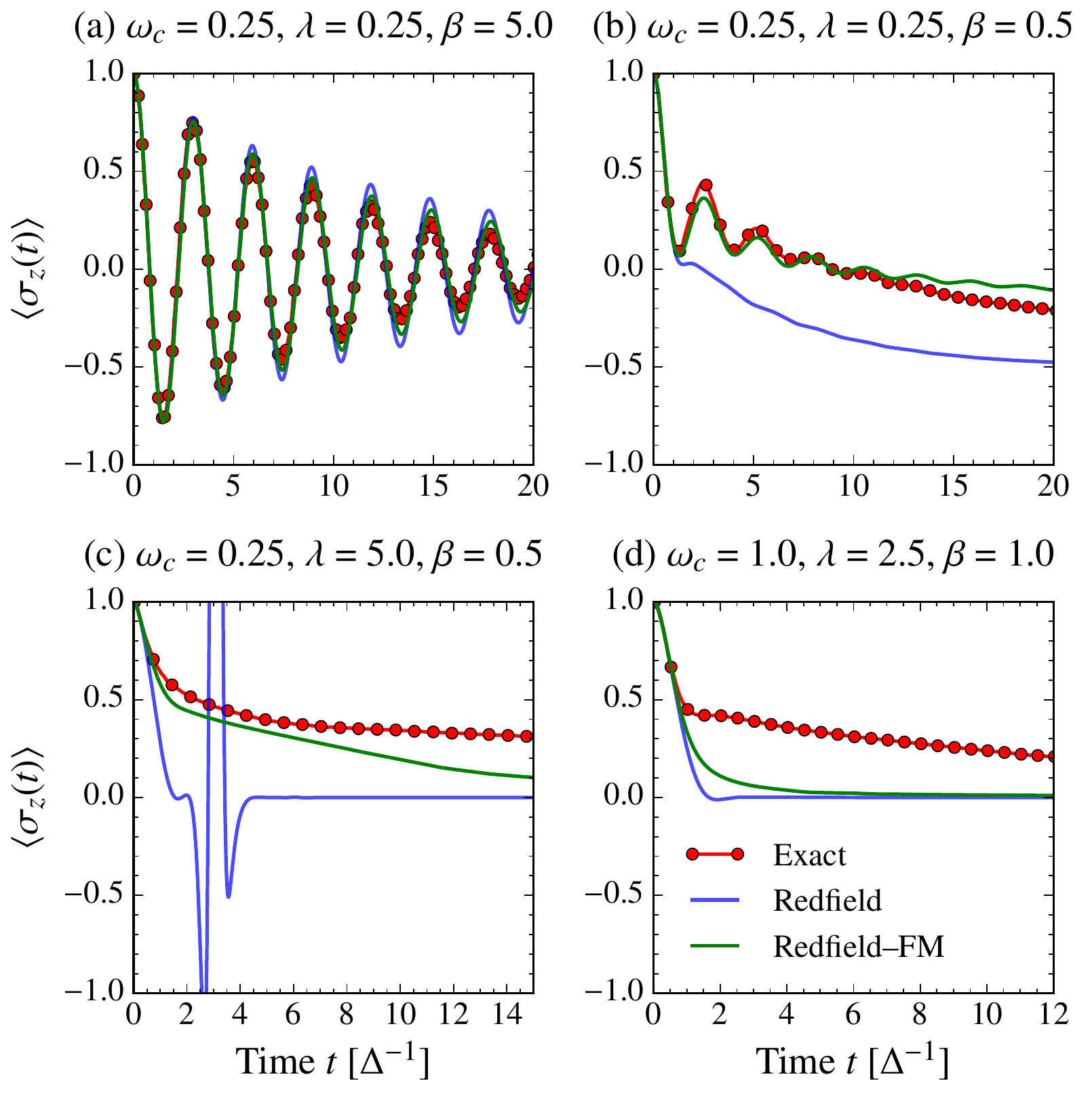} 
\caption{Results from Redfield--FM approach compared with standard time--local
Redfield theory and exact numerics.   $\omega^* = \mathrm{max}[\omega_c,
\frac{\omega_R}{4}]$ and $\omega_R = 2\sqrt{\Delta^2 + \epsilon^2}$. All units
are scaled by the electronic coupling $\Delta$. All panels correspond to
unbiased cases ($\varepsilon = 0$), except for panel (b), where $\varepsilon =
1.0$.  Other parameters are stated in the panels.}
\end{figure}

As mentioned in Sec.~\ref{Sec:Theory:Redfield}, the validity of Redfield theory
is limited to the small $\eta$ regime.  Fig.~2(a) shows the results for a slow
bath ($\omega_c = 0.25$) with small reorganization energy ($\lambda = 0.25$) at
low temperature ($\beta = 5.0$); here and in the following, all energies are in
units of $\Delta$.  In spite of the slow bath, the dimensionless applicability
parameter is only slightly larger than unity ($\eta = 1.6$) suggesting that
Redfield theory should be reasonably accurate, in agreement with the numerical
results.  The Redfield--FM method provides an even better estimate of the
dynamics, almost quantitatively correcting the already accurate Redfield
dynamics.  

Fig.~2(b) considers a biased system ($\varepsilon = 1.0$) with the same
parameters, except at much 
higher temperature ($\beta = 0.5$), yielding an applicability parameter which
is now significantly larger than unity ($\eta = 16$).  Here, it is evident that
the Redfield dynamics relax far too quickly, suppressing the coherence and
missing the slower relaxation process revealed by the exact dynamics.  The
improvement afforded by the Redfield--FM method compared to standard Redfield
theory is clear.  In particular, the Redfield--FM approach accurately
reproduces the short-- to intermediate--time dynamics, the frequency 
of the oscillations, and the initial rate of decoherence.  The terminal decay
rate is slightly underestimated due to the mode--freezing
approximation, as discussed in Section II.B.  Yet despite these shortcomings,
the improvement derived from a simple scheme like the Redfield--FM approach
with the numerical complexity of the original Redfield theory is noteworthy.   

For cases exemplified by Fig.~2(c), serious problems such as the violation of
the positivity of the RDM dynamics can occur within standard (non--secular) Redfield theory.
Figure~2(c) corresponds to a slow bath ($\omega_c = 0.25$) and a large
reorganization energy ($\lambda = 5.0$) again at high temperature ($\beta = 0.5$), 
for which the applicability parameter
is very large ($\eta = 320$).   Despite the evident failure of the Redfield
equations to even maintain positivity, the Redfield--FM method is able to
correct the positivity issue and almost quantitatively reproduce the two--step
relaxation process in the exact dynamics up to intermediate times.  

It is possible to understand the surprising success presented in Fig.~2(c) in
the context of the analysis of Section II.B.  Using the definitions given in
that section, the effective parameters for the Redfield equation
are $\lambda_{\mathrm{fast}} = 2.5$, $\omega_c = 0.5$, yielding $\eta = 40$.  
Although $\eta \gg 1$, the reduction by an order of magnitude from the initial 
value, $\eta = 320$, is sizeable, and likely responsible for solving the positivity problem
evident in the bare Redfield dynamics.  The reproduction of the two--step
relaxation process is a direct result of the trapping effect that arises from
freezing a large portion of the low--frequency bath in the presence of large
coupling.  This example indicates that the trapping effect can partially
reproduce slow relaxation dynamics associated with strong system--bath interactions. 
    
Figure~2(d) shows the regime of intermediate bath speeds ($\omega_c = 1$),
large reorganization energy ($\lambda = 2.5$), and intermediate temperature
($\beta = 1$).  In contrast to Fig.~1(c), the Redfield--FM
method is not capable of significantly improving the Redfield dynamics in this
regime, missing the two--step relaxation process visible in the exact dynamics.
In light of the previous case, it is evident that the slowing down of the RDM
dynamics can be caused by freezing a large portion of the strongly coupled
modes, an effect which is absent in this case. In this case, $\lambda_{\mathrm{fast}} =
2.1$ and $\lambda_{\mathrm{slow}} = 0.4$, which indicates that most of the
reorganization energy is included already in the high--frequency portion of the
bath.  In such cases, the Redfield--FM method will yield results that are
similar to bare Redfield theory.  
    	
\begin{figure}
\includegraphics[width=8.5cm]{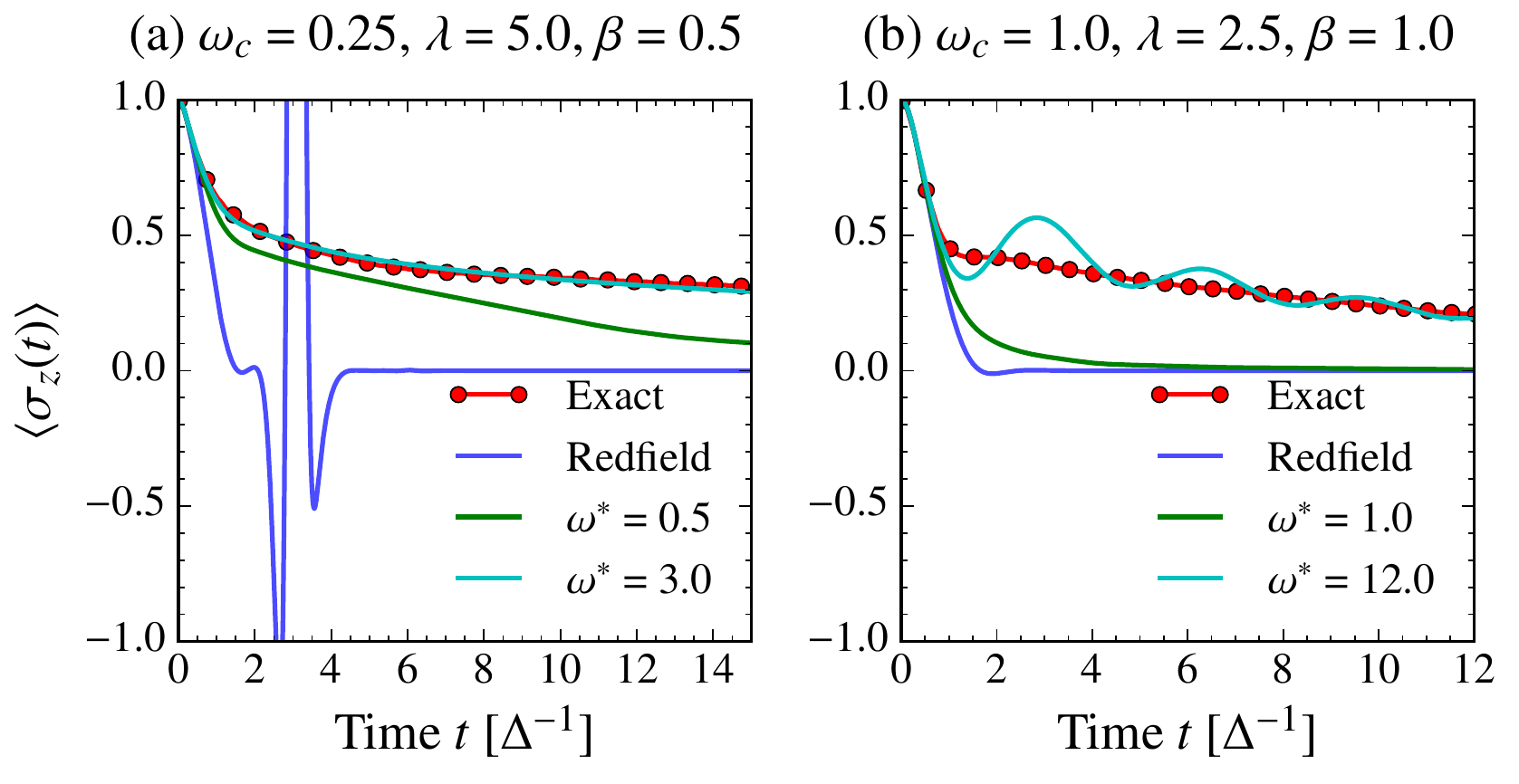} 
\caption{Redfield--FM results for $\omega^* = \mathrm{max}[\omega_c,
\frac{\omega_R}{4}]$ and an ``optimized'' value for $\omega^*$. Both cases
considered here correspond to $\varepsilon = 0$ and all units are scaled by the
electronic coupling, $\Delta$. Note that the set of parameters for panels (a)
and (b) in this figure correspond to the set of parameters in Fig.~2, panels
(c) and (d), respectively.}
\end{figure}

We now address the dependence of the dynamics on the choice of $\omega^*$.
Eschewing the simple criteria for choosing $\omega^*$ presented above, one may
ask how closely the Redfield--FM dynamics can be made to agree with exact
dynamics when $\omega^*$ is allowed to vary.  To address this question, we
include two extreme cases in Fig.~3.  First, Fig.~3(a), which corresponds to the same
parameters as those of Fig.~2(c), shows that optimization of $\omega^*$ can
result in \textit{quantitative} agreement between the Redfield--FM result and
the exact dynamics. Such agreement may be understood as the result of
fortuitous cooperation between strongly dissipative Redfield dynamics that damp
the frozen mode--generated oscillations and the trapping effect from the
mode--freezing approximation that prevents immediate relaxation to the
equilibrium population.  Conversely, Fig.~3(b), which corresponds to the
parameters in Fig.~2(d), is an example of when perfect agreement is
impossible. Clearly, attempts at optimizing $\omega^*$ result in better
agreement of the two--step relaxation process at the cost of long--lived
oscillations, a direct result of including a large fraction of modes into the
slow part of the bath.  In freezing a sufficiently large portion of the
reservoir to reproduce the trapping effect, $\lambda_{\mathrm{fast}}$ is reduced to the
point where the Redfield dynamics are no longer sufficiently dissipative to
damp the frozen mode--generated oscillations.  In addition, $\lambda_{\mathrm{slow}} =
\pi^{-1}\int_0^{\infty} d\omega J_{\mathrm{slow}}(\omega)/\omega$ is also not large
enough to ensure that the oscillations dephase sufficiently rapidly.  Overall,
it is clear that although it may be possible to optimize the results, the
simple initial criteria presented represent a robust approach to frozen mode
dynamics that essentially always yields results that are as good or better than
bare Redfield dynamics without a significantly increased computational cost.        

\subsection{Relaxing the Mode--Freezing Approximation: A Dynamical Hybrid Redfield Method}
\label{Sec:Results:Hybrid}

On first inspection, the mode--freezing approximation appears extreme.  To
thoroughly assess its effect, we develop a dynamical hybrid method in which we
evolve the previously frozen low--frequency modes in
$J_{\mathrm{slow}}(\omega)$ via classical Ehrenfest dynamics.  The derivation and the
implementation details of this approach may be found in Appendix \ref{App:RDMHybrid}.  

This hybrid--Redfield method is similar in spirit to the successful
hybrid--NIBA developed and implemented in Ref.~\onlinecite{Berkelbach2012}.  
Evolution of the low--frequency modes using
Ehrenfest dynamics in such hybrid approaches only requires two assumptions: (i)
that $\Gamma(t) \approx \rho_{\mathrm{slow}}(t)\rho_{\mathrm{sys+fast}}(t)$,
and (ii) that the motion of the low--frequency modes is well--captured by
classical mechanics.  For such a factorization to be valid, the reorganization
energy due to the low--frequency bath needs to be small, i.e.,
$\lambda_{\mathrm{slow}} = \pi^{-1}\int_0^{\infty} d\omega\
J_{\mathrm{slow}}(\omega)/\omega < \Delta$.  The applicability of classical
dynamics relies on the low energies of the reservoir
modes and sufficiently high temperatures that help suppress quantum
effects.\cite{Stock1995a, Berkelbach2012}  However, even when the Ehrenfest
approximation is valid, problems may arise.  Most prominent among these is that
the final populations approach those of the infinite--temperature
limit.\cite{Stock1995a}  

In contrast to the Hamiltonian derived under the mode--freezing approximation
in Eq.~(\ref{Eq:Hsc}), the modified Hamiltonian that needs to be treated via
the Redfield equation in the hybrid--Redfield method is time--dependent,
    \begin{equation}
    \begin{split}
    H''(t) &= [\varepsilon + \lambda^{cl}(t)]\sigma_z + \Delta\sigma_x\\
    &\hspace{1em} + \sigma_z \sum_{k\in \text{fast}}c_{k}Q_{k} +\frac{1}{2}\sum_{k}\Big[P_{k}^2 + \omega_{k}^2Q_{k}^2 \Big],
    \end{split}
    \end{equation}
where the disorder due to the low frequency bath is no longer static as it is
in the Redfield--FM method, but rather dynamic, namely $\lambda^{cl}(t) =\sum_k
c_{k}Q_{k}(t)$.

Since the system part of this Hamiltonian is nondiagonal and time--dependent,
evolution with respect to the system Hamiltonian requires diagonalization at
every time--step, significantly increasing the computational cost associated
with the method proposed here.  The need to evolve the low--frequency bath also
adds to the computational cost of the approach.  Importantly, under the
mode--freezing approximation, we circumvent these costly requirements.  This
means that, aside from the trivial cost of parallelization for the ensemble
averaging over the slow bath, the Redfield--FM method scales as gracefully with
system size as the original Redfield equation. 
    
For completeness, we remark that the nonequilibrium population dynamics under
the hybrid--Redfield approximation now take the form
    \begin{equation}
    \langle \sigma_z(t)\rangle  \approx \int d\mathbf{P}d\mathbf{Q}\ \rho_{\mathrm{slow}}(\mathbf{P}, \mathbf{Q}, t) \Tr_{\mathrm{sys+fast}}[\rho_{\mathrm{sys+fast}}(t)\sigma_z].
    \end{equation}
    
Fig.~4 shows two sets of parameters for which the hybrid--Redfield scheme
yields results that illustrate the issues at play in comparing the
hybrid--Redfield approach to the Redfield--FM method. Extensive testing of the
hybrid method suggests that an approximately optimal form for the splitting
frequency can be taken as 
\begin{equation}
\omega^*_{hy} = \omega_R\lambda/\omega_c.
\end{equation}  
Physically, this form encodes the interplay between the Redfield and
Ehrenfest methods, favoring a larger portion of the modes to be treated
classically with increasing Rabi frequency $\omega_R$, which is a measure of
how rapidly the electronic system evolves.  Furthermore, this form of
$\omega^*$ ensures that in the limit of small $\lambda$ and large $\omega_c$,
the hybrid method correctly reproduces the more appropriate Redfield dynamics,
whereas in the limit of large $\lambda$ and small $\omega_c$, it reproduces the
Ehrenfest results.  It is expected that generally nontrivial results may be
obtained from this method for cases where $\omega^* \sim \omega_c$, as is the
case for the choice of $\omega^*$ used in the Redfield--FM approach.  

\begin{figure}
\includegraphics[width=8.5cm]{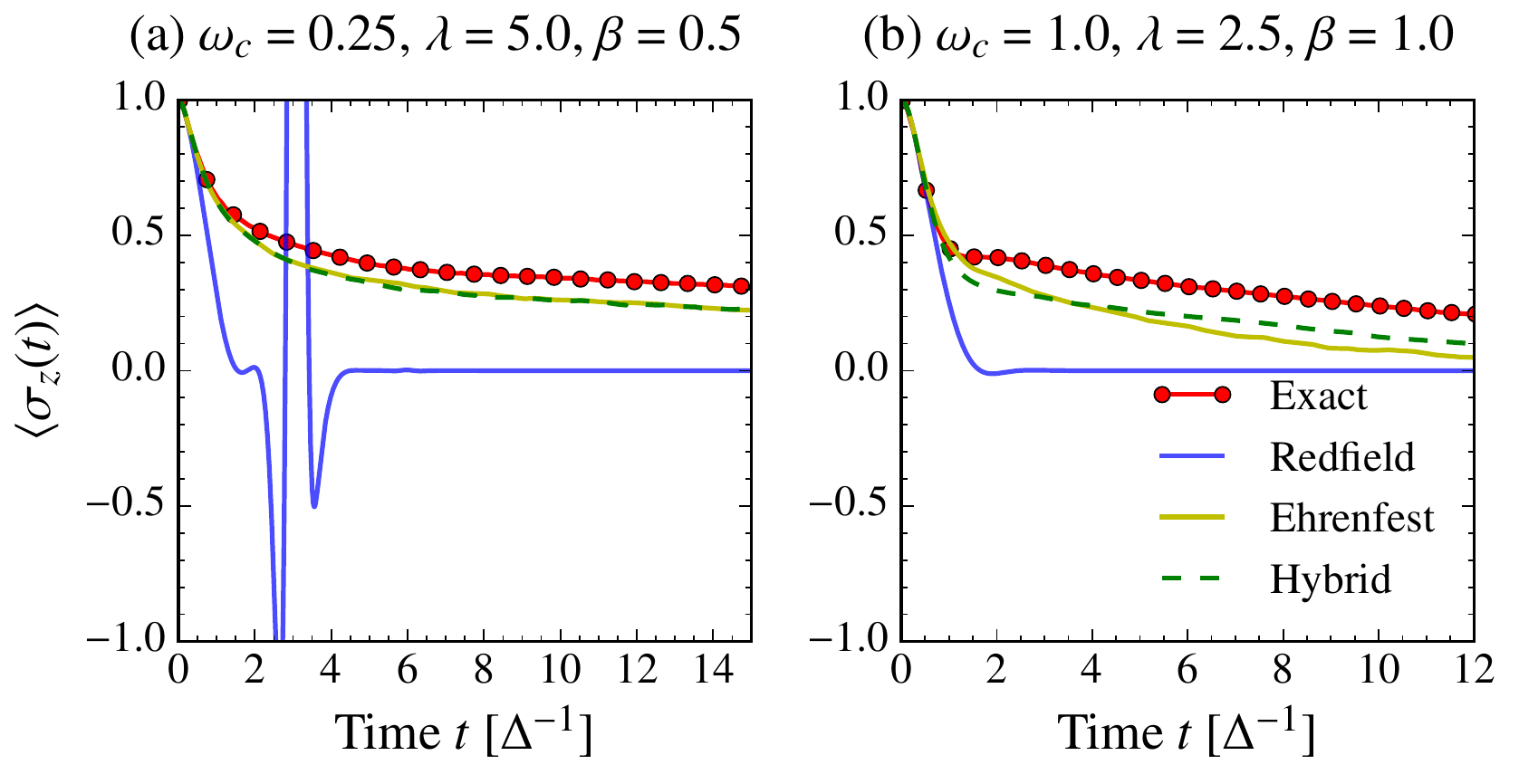} 
\caption{Hybrid--Redfield results for $\omega^*_{hy} = \omega_R\frac{\lambda}{\omega_c}$.  Both cases considered here correspond to $\varepsilon = 0$ and all units are scaled by the electronic
coupling, $\Delta$. Similar to Fig.~3, the set of parameters for panels (a) and (b) in this figure correspond to the set of parameters in Fig.~2, panels (c) and (d), respectively.}
\end{figure}

Fig.~4(a) corresponds to the parameters in Fig.~2(c) and illustrates that, by
means of the suggested form for $\omega^*$,  hybrid--Redfield automatically
tunes itself to yield nearly optimal results achievable from the two parent
methods. This example, for which $\omega^*_{hy} \gg \omega_c$, illustrates that
the hybrid--Redfield method trivially reproduces the Ehrenfest result when it
is appropriate.  \textit{It is noteworthy that the Redfield--FM method obtains
similar agreement at a much lower computational cost} without evolving the
reservoir modes, indicating that dynamic treatment of these modes may not be
generally necessary. Indeed, it is rather remarkable that the Redfield--FM
approach basically recapitulates the Ehrenfest results even though no Ehrenfest
dynamics are used.   

Fig.~4(b) shows the analogue of Fig.~2(d), where the Redfield--FM method fails
to correct the Redfield dynamics.  In contrast, the hybrid--Redfield results
are in very good agreement with the exact dynamics.  Indeed, the hybrid method
is able to qualitatively and almost quantitatively reproduce the shape of the
two--step relaxation process evident in the exact dynamics, an effect that both
Ehrenfest and Redfield dynamics independently miss.  

As the above considerations indicate, there are cases where the
dynamical hybrid--Redifeld method can provide a substantial improvement over the
Redfield--FM method, albeit at a much higher computational cost. In most
regions of parameter space we have studied, however, we find that
hybrid--Redfield theory offers little accuracy gain over the Redfield--FM
approach.  Thus, the benefits of the hybrid--Redfield approach compared to the
Redfield--FM method do not justify its use when accuracy and cost are factored
together. 

\section{Conclusions}
\label{Sec:Conclusions}

In this work, we have presented a new scheme for simulating dynamics in quantum
dissipative systems.  Our approach, which we call the Redfield--FM method,
recognizes that standard Redfield theory becomes inaccurate for slow bath
degrees of freedom.  By partitioning the bath into high-- and low--frequency
components, we propose solving the Redfield equations for the high--frequency
partition in the statically disordered field of the low--frequency components.
Such an approach may greatly increase the accuracy of Redfield theory in highly
non--Markovian regimes at
essentially the same computational cost.  In addition, we find that this simple
approach can fundamentally cure positivity problems associated with standard
non--secular Redfield theory.  We have further discussed a scheme (the hybrid--Redfield
approach) whereby the previously frozen degrees of freedom are instead evolved
with classical Ehrenfest dynamics.  While this method can improve upon the dynamics as
described by the Redfield--FM approach, the increase in accuracy is incremental
and comes at a significantly larger computational cost.  Overall, while the
Redfield--FM method does not cure all of the ills of Redfield theory, it does
provide a simple and efficient framework for improving its accuracy and range
of validity, especially for sluggish bath degrees of freedom such as those
implicated in biological energy transfer.

\begin{acknowledgements}
DRR and AMC acknowledge support from NSF CHE-1464802.  TCB was supported by the Princeton Center for Theoretical Science.
\end{acknowledgements}

\appendix

\section{Derivation of Redfield equations}
\label{App:Redfield}

Here, for completeness, we review the derivation of the Redfield equations.
For a more detailed discussion of Redfield theory, we refer the reader to
Refs.~\onlinecite{BreuerPetruccione} and \onlinecite{Chang1993}.

In the following development we utilize a projection operator technique to
derive an equation of motion for the reduced density matrix (RDM) of the
system, defined as $\rho(t) = \Tr_{\mathrm{bath}}[\Gamma(t)]$, where $\Gamma(t)
= e^{-iHt}\Gamma(0)e^{iHt}$ and $\Gamma(0)$ is the initial density matrix of
the full system and bath.  Moreover, we assume that the initial condition for
the (total) density matrix contains no system--bath correlation, such that
$\Gamma(0) = \rho(0)\rho_{\mathrm{bath}}$, where $\rho(0)$ is an arbitrary
Hermitian system operator, $\rho_{\mathrm{bath}} = e^{-\beta
H_{\mathrm{bath}}}/Z$, $Z = \Tr_{\mathrm{bath}}[e^{-\beta H_{\mathrm{bath}}}]$
and $\beta = 1/k_BT$ is the thermal energy.  Treatment of general initial
conditions is also possible via the projection operator technique at the
expense of the introduction of additional inhomogeneous terms in
Eq.~(\ref{Eq:SchroDensityOperatorRedfield}).\cite{Zwanzig1961, Jang2002,
Shibata1980}  In the following, we ignore initial correlations, but note that
their inclusion in the present framework is straightforward.

We start from the Liouville equation for the full density matrix in the
interaction picture where the total Hamiltonian is divided into a zeroth order
part and an interaction part, $H = H_0 + H_1$, such that
    \begin{equation}\label{Eq:LiouvilleEq}
    \frac{d}{dt}\Gamma_I(t) = -i\mathcal{L}_I(t)\Gamma_I(t),
    \end{equation}
$\Gamma_I(t) = e^{iH_0t}\Gamma(t)e^{-iH_0t}$ and $\mathcal{L}_I(t) =
[e^{-iH_0t}H_1e^{iH_0t}, ...]$.  To obtain the dynamics of the RDM, we define a
projection operator of form $\mathcal{P} \equiv
\rho_{\mathrm{bath}}\Tr_{\mathrm{bath}}[...]$ with $Q \equiv 1 - \mathcal{P}$.
We note that action of $\mathcal{P}$ on the full density matrix followed by
trace over the bath results in the RDM in the interaction picture, $\rho_I(t) =
\Tr_{\mathrm{bath}}[\mathcal{P}\Gamma_I(t)]$.  Using these definitions, we
obtain the following exact equations of motion,     
    \begin{align}
    \frac{d}{dt} \mathcal{P}\Gamma(t) &= -i \mathcal{P}\mathcal{L}_I(t)(\mathcal{P} + \mathcal{Q}) \Gamma(t)\\ 
    \frac{d}{dt} \mathcal{Q}\Gamma(t) &= -i \mathcal{Q}\mathcal{L}_I(t)(\mathcal{P} + \mathcal{Q}) \Gamma(t). \label{Eq:ProjDynamics}
    \end{align}
Formal integration of Eq.~(\ref{Eq:ProjDynamics}) yields
    \begin{equation}\label{Eq:QGamma}
    \mathcal{Q}\Gamma_I(t) =  - i\int_0^t d\tau\ g(t, \tau) \mathcal{Q} \mathcal{L}_I(t) \mathcal{P} \Gamma_I(\tau),
    \end{equation}
where $g(t, \tau) = \exp_+[-i\int_\tau^t ds\ \mathcal{Q}\mathcal{L}_I(s)]$ and
the time ordering ($+$) implies that time arguments increase from right to
left. Substitution of this expression in Eq.~(\ref{Eq:LiouvilleEq}) results
in the Nakajima--Zwanzig equation,\cite{Zwanzig1961, Jang2002} which is
expressed in terms of the time convolution of a memory term with $\rho_I(t)$ at
earlier times as 
    \begin{equation}\label{Eq:TimeNonLocal}
    \frac{d}{dt}\rho_{I}(t) =  - \int_0^t d\tau\ K(t - \tau)\rho_I(\tau), 
    \end{equation}
where $K(t - \tau) = \Tr_{\mathrm{bath}}[\mathcal{L}_I(t)g(t,
\tau)\mathcal{Q}\mathcal{L}_I(\tau)\rho_{\mathrm{bath}}]$ is the
(time--nonlocal) memory function.

If, instead, we use the formal solution of Eq.~(\ref{Eq:LiouvilleEq}) to evolve
$\Gamma_I(t)$ backwards in time to an earlier time $\tau$, we obtain
    \begin{equation}
    \Gamma_I(\tau) = G(t, \tau)\Gamma_I(t), 
    \end{equation}
where $G(t, \tau) = \exp_-[i\int_\tau^t ds \mathcal{L}_I(s)]$ and the time
ordering ($-$) requires that time arguments increase from left to right.
Replacing this expression in Eq.~(\ref{Eq:QGamma}) and solving for
$\mathcal{Q}\Gamma_I(t)$ yields
    \begin{equation}\label{Eq:QGamma2}
    \mathcal{Q}\Gamma_I(t) =  [1 - \Sigma(t)]^{-1}\Sigma(t)\mathcal{P}\Gamma_I(t), 
    \end{equation}
where
    \begin{equation}
    \Sigma(t) = -i\int_0^td\tau\ g(t, \tau)\mathcal{Q}\mathcal{L}(\tau)\mathcal{P}G(t, \tau).
    \end{equation}
We note that a crucial requirement for the validity of this derivation is the
existence of $[1 - \Sigma(t)]^{-1}$.  

Substitution of Eq.~(\ref{Eq:QGamma2}) into Eq.~(\ref{Eq:LiouvilleEq}) and
subsequent trace over the bath degrees of freedom results in the following
time--local equation of motion for the RDM,\cite{Shibata1980} 
    \begin{equation}\label{Eq:TimeLocal}
    \frac{d}{dt}\rho_{I}(t) =  R(t)\rho(t), 
    \end{equation}
where $R(t) = - i \Tr_{\mathrm{bath}}[\mathcal{L}_I(t)[1 -
\Sigma(t)]^{-1}\rho_{\mathrm{bath}}]$ is the (time--local) rate function.

The expression for the dynamical evolution in either the time--nonlocal
(Eq.~(\ref{Eq:TimeNonLocal})) or time--local (Eq.~(\ref{Eq:TimeLocal})) form is
exact but prohibitively difficult to evaluate without resorting to
approximation schemes, such as truncated generalized cumulant expansions.
Perturbative expansion to second order in the system--bath coupling (where $H_1
= V$ from Eq.~(\ref{Eq:SBCoupling})) results in a non--Markovian generalization of
the Redfield theory.  Alternatively, one may derive both forms of the Redfield
equations via resummations of differently time--ordered
cumulants.\cite{Yoon1975, Mukamel1978}  These derivations explicitly show that
both forms of generalized Redfield theory account for non--Markovian behavior
and have similar applicability requirements.\cite{Mukamel1978, Breuer1999,
Coalson2004}  Specifically, since Redfield theory is tantamount to
second--order perturbation theory in the system--bath coupling, truncation at
low order is only accurate for $\eta < 1$, where $\eta =
\mathrm{max}[\frac{2\lambda}{\beta \omega_c^2}, \frac{2\lambda}{\pi \omega_c}]$
is the validity parameter introduced in Sec.~\ref{Sec:Theory:Redfield}.
Despite this restriction, the Redfield equations have been shown to  perform
surprisingly well, often beyond the small--$\lambda$ and large--$\omega_c$
regimes.\cite{Yang2002, Palenberg2001}  Nevertheless, for inappropriate regions
of parameter space, severe problems can arise, such as violation of positivity
in the reduced density matrix.\cite{Nitzan}

\section{Markovian Redfield theory with Frozen Modes}
\label{App:Markovian}

We wish to consider the performance of the frozen modes method for a strictly
Markovian version of Redfield theory, i.e. with a rate tensor $R =
R(t\rightarrow \infty)$.  In this limit, the time integrals become
Fourier--Laplace transforms, such that the Redfield tensor elements can be
expressed algebraically in terms of the spectral density $J(\omega)$ evaluated
at energy differences $\hbar\omega_{ij}\equiv
(E_i-E_j)$.\cite{Pollard1996,Ishizaki2009a}  
More specifically, we are interested in the dephasing terms of the Redfield
tensor, which in general contain an elastic contribution
\begin{equation}
R_{ijij} \sim g_{ij}^2\ J(\omega=0+)\ n_{BE}(\omega=0+).
\end{equation}
At low frequencies, the Bose--Einstein distribution, $n_{BE}(\omega) \sim kT/\omega$, such that for a spectral density
of the form $J(\omega) \sim \omega^{s}$, we find
\begin{equation}
R_{ijij} \sim kT \omega^{s-1} \Big|_{\omega=0}.
\end{equation}
For all `super--Ohmic' spectral densities with $s>1$, this elastic contribution
to the dephasing rate vanishes.  However, for an Ohmic spectral density with $s=1$
there is a pure dephasing rate which vanishes only at $T=0$.  This contribution
to the dephasing rate in the system's eigenbasis can significantly affect
both the population and coherence dynamics in the original basis of the problem.

We now return to the idea of a frozen modes variant of Markovian Redfield
theory.  Consider specifically an Ohmic spectral density with any non--zero
splitting frequency $\omega^*$. After partitioning, the fast spectral density
has the low--frequency behavior $J_{\mathrm{fast}}(\omega) \sim \omega^s$
with $s>1$, which yields no elastic contribution to the dephasing rate.  For this
reason, a frozen modes version of Markovian Redfield theory does not reduce to the
Redfield limit until the singular point $\omega^* = 0$.  Instead, as $\omega^*\rightarrow
0$, the result approach a Redfield result which neglects the Ohmic pure dephasing
rate.  We emphasize that the time--dependent variants of Redfield theory are not
signficantly affected by this problem until very long times, and that all methods
are only affected for strictly Ohmic spectral densities.

We propose a very simple solution to this pathological
behavior in Markovian Redfield theory, by modifying the fast spectral density via
    \begin{equation}
    J_{\mathrm{fast}}(\omega) = [1-S(\omega, \omega^*)]J(\omega) 
        +\  W(\omega,\epsilon) J(\omega), 
    \end{equation}
where $W(\omega,\epsilon)$ is a rectangular window function centered
at the origin with width $\epsilon$, and $\epsilon$ should be chosen very
small.  In this way, the `fast' part of the bath will always produce a pure
dephasing rate for arbitrary splitting frequency $\omega^*$.  Thus, the
Markovian Redfield--FM dynamics will smoothly interpolate towards the standard
Markovian Redfield result as $\omega^*$ approaches zero.  In Fig.~5, we compare
the results of standard Markovian Redfield, Markovian Redfield--FM without this
dephasing correction, and Markovian Redfield--FM with the correction.  Results
are presented for the model excitonic dimer discussed by Ishizaki and
Fleming.\cite{Ishizaki2009b} Importantly, we find that this correction
typically improves the results of the Markovian Redfield--FM variant, quite
significantly in cases of strong system--bath coupling.

\begin{figure}
\includegraphics[width=8.5cm]{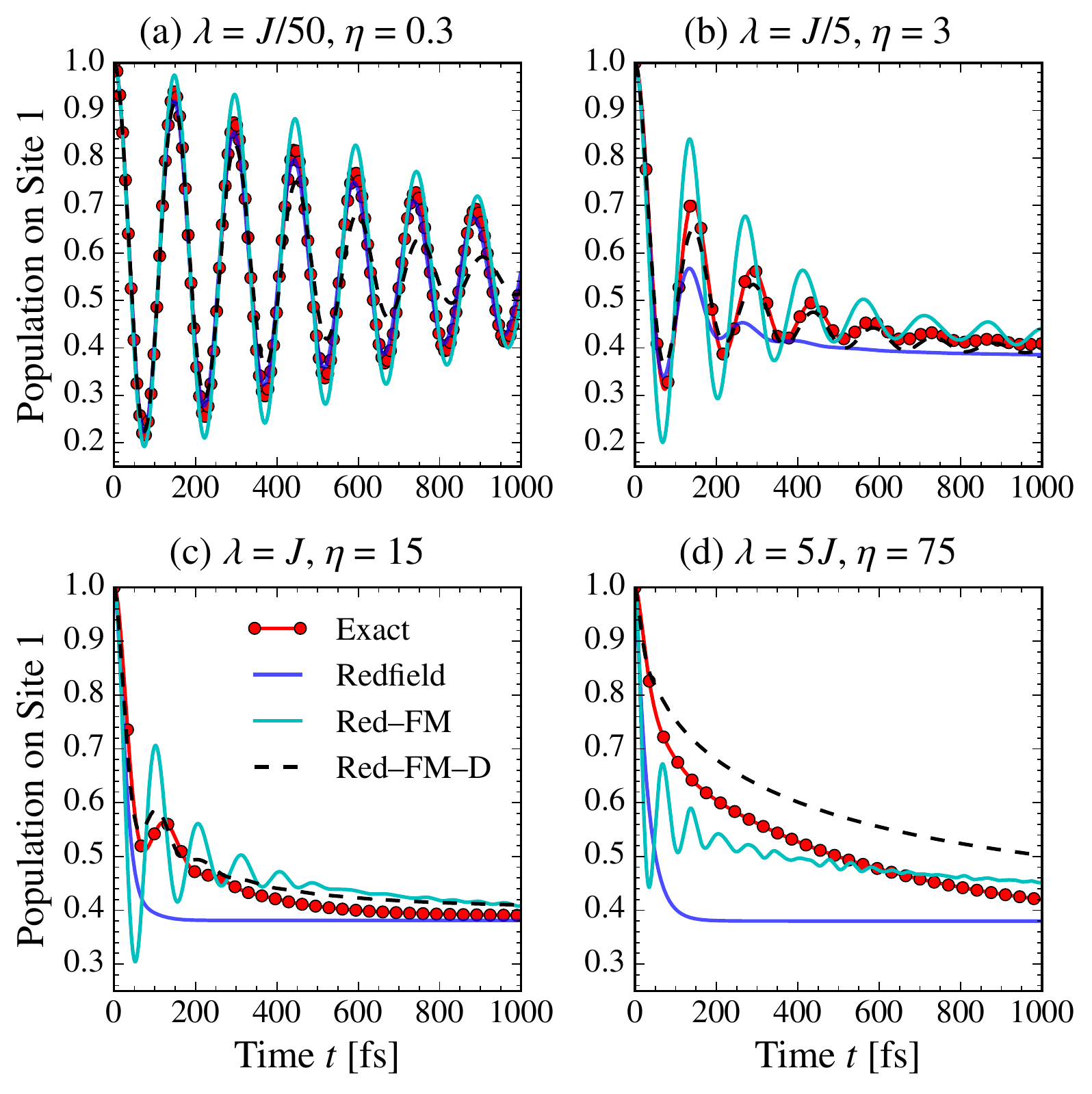} 
\caption{
Comparison of numerically exact (HEOM) population dynamics to the
results of standard Markovian Redfield theory, a straightforward
variant of Markovian Redfield theory with frozen modes (Red--FM), and a dephasing--corrected variant as discussed in the text (Red--FM--D).  The system--bath Hamiltonian
is that of Ref.~\onlinecite{Ishizaki2009b} with $\varepsilon = 100$ cm$^{-1}$,
$J = 100$ cm$^{-1}$, $\omega_c^{-1} = 100$ fs, and $T = 300$ K.
}
\end{figure}

\section{RDM Hybrid Method}
\label{App:RDMHybrid}

Here we relax the mode--freezing approximation by deriving a fully hybrid
method that separates the complete system into a slow part consisting of the
low--frequency component of the bath, and a rapidly--evolving part that
includes both the electronic system and the high--frequency portion of the
phonon bath.  In this hybrid scheme, the slow part is treated
quasi--classically, while the fast part is treated at the level of Redfield
theory.  Overall, the fast (slow) component of the system evolves in the mean
field of the slow (fast) one.  

Other hybrid approaches that combine classical and quantum dynamics include the
self--consistent hybrid method of Wang and coworkers,\cite{Wang2001,Thoss2001} which
yields numerically exact dynamics, and the approximate hybrid--NIBA approach of
Refs.~\onlinecite{Berkelbach2012,Berkelbach2012a}  In the former method,
$\omega^*$, which is the energy scale that determines the splitting of the bath
into slow and fast parts, is strictly a convergence parameter.  In the latter,
$\omega^*$ is an empirically determined adjustable parameter.  As an
approximate method, the hybrid--Redfield scheme derived here is akin to the
hybrid--NIBA method.  For a more detailed discussion of the hybrid RDM method,
we refer the reader to Ref.\ \onlinecite{Berkelbach2012}.

As in the Redfield--FM approach, we make the approximation that $\Gamma(t)
\approx \rho_{\mathrm{slow}}\rho_{\mathrm{sys+fast}}(t)$, 
where $\rho_{\mathrm{sys+fast}}(t)$ is the density matrix for
the system and fast bath degrees of freedom and $\rho_{\mathrm{slow}}(t)$ is the 
density matrix for the slow bath degrees of freedom.  
The system and fast bath modes obey the following effective Liouville
equation,
    \begin{equation}
    \frac{d\rho_{\mathrm{sys+fast}}(t)}{dt} = -i[H''(t), \rho_{\mathrm{sys+fast}}(t)],
    \end{equation}
where
    \begin{equation} \label{Eq:Hsc11}
    \begin{split}
    H''(t) &=  [\varepsilon + \lambda^{cl}(t)]\sigma_z + \Delta\sigma_x\\
    &\hspace{1em} + \sigma_z\sum_{k \in \text{fast}}c_{k}Q_{k}+\frac{1}{2}\sum_{k}\Big[P_{k}^2 + \omega_{k}^2Q_{k}^2 \Big] ,
    \end{split}
    \end{equation}
and $\lambda^{cl}$ is a dynamically fluctuating bias, $\lambda^{cl}(t)
=\sum_{k\in \text{slow}} c_{k}Q_{k}(t)$.

A classical treatment of the reservoir leads to the following equations of motion,
    \begin{equation}
    \frac{dQ_{k}}{dt} = P_{k},
    \end{equation}
and
    \begin{equation}
    \begin{split}
    \frac{dP_{k}}{dt} &= -\omega_{k}^2Q_{k}  - c_{k}\tilde{\sigma}_z(t).
    \end{split}
    \end{equation}
    Employing the Ehrenfest approach demands that each part of the system
    evolves in the mean field of the other.  For the quantum portion, the
    classical mean field consists of the time--dependent contribution to the
    bias, $\lambda^{cl}(t)$.  For the classical portion, the force term
    $-c_k\tilde{\sigma}_z(t) = -c_k\Tr_{\mathrm{sys+fast}}[\sigma_z\rho_{\mathrm{sys+fast}}(t)]$
    in the equations of motion embodies the mean--field `back--reaction.'  This force
    term moves the classical oscillators from the ground state minima to the
    displaced minima associated with the excited state. 

Using Eqs.~(\ref{Eq:SchroDensityOperatorRedfield}) and (\ref{Eq:Hsc11}), the
time--local  Redfield equation takes the form
    \begin{equation}\label{Eq:SchroDensityOperator}
    \begin{split}
    \frac{d}{dt} \rho(t) &= -i [H''_{\mathrm{sys}}(t), \rho(t)]\\
    &\hspace{1em} - \int_0^{t} ds\ \Big\{ C(s)[\sigma_z, \sigma_z(-s)\rho(t)]\\
    &\hspace{5em} - C^*(s)[\sigma_z, \rho(t)\sigma_z(-s)]\Big\},
    \end{split}
    \end{equation}
where $\sigma_z(t) = U_0^{\dagger}(t)\sigma_zU_{0}(t)$, $U_0(t) =
\exp[-i\int_0^t d\tau\ H_{\mathrm{sys}}''(\tau)]$, and $H_{\mathrm{sys}}''(t) = [\varepsilon +
\lambda^{cl}(t)]\sigma_z + \Delta\sigma_x +\sum_{k\in\text{slow}}[P_{k}^2 +
\omega_{k}^2Q_{k}^2 ]/2$.  The bath correlation, as in the case of the
Redfield--FM method, takes the following form,
    \begin{equation}
    \begin{split}
    C(t) &= \frac{1}{\pi}\int_0^{\infty}d\omega\ J_{\mathrm{fast}}(\omega)\\
    &\hspace{4em} \times [ \coth(\beta\omega/2)\cos(\omega t) - i\sin(\omega t) ].
    \end{split}
    \end{equation}

To obtain the results shown in Fig.~3, trajectories corresponding to a set of
initial conditions sampled from the Wigner distribution\cite{Hillery1984} were
calculated via a second--order Runge--Kutta scheme, using a step size of
$\delta t = 0.01\Delta^{-1}$. As required by the Runge--Kutta procedure,
$\tilde{\sigma}_z(t)$ was kept constant during the evolution of the bath while
$\lambda^{cl}(t)$ was kept constant during the evolution of the system.
Explicitly, over a half--time step, the equations for the bath become
    \begin{equation}
    \begin{split}
    Q_{k}\left(t + \frac{\delta t}{2} \right) &= \alpha_{k}(t)\cos\left( \frac{\omega_{k} \delta t}{2} \right) - \frac{c_{k}}{\omega_{k}^2}\tilde{\sigma_z}(t)\\
    &\hspace{1em} + \frac{P_{k}(t)}{\omega_{k}}\cos\left(\frac{\omega_{k} \delta t}{2}\right) ,
    \end{split}
    \end{equation}
and 

    \begin{equation}
    \begin{split}
    P_{k}\left(t + \frac{\delta t}{2} \right) &= P_{k}(t)\cos\left(\frac{\omega_{k} \delta t}{2} \right)+ \omega_{k}\alpha_{k}(t)\sin\left(\frac{\omega_{k} \delta t}{2}\right),
    \end{split}
    \end{equation}
where
    \begin{equation}
    \alpha_{k}(t) =  Q_{k}(t) + \frac{c_{k}}{\omega_{k}^2}\tilde{\sigma_z}(t).
    \end{equation}

In the hybrid--NIBA method of Ref.~\onlinecite{Berkelbach2012}
the zeroth--order propagator necessary to evolve the perturbation in the
interaction picture, $U_0(t) = \exp[-i\int_0^{\infty}d\tau H_{\mathrm{sys}}''(\tau)]$,
was simple to calculate since $H_{\mathrm{sys}}''$ was diagonal. In contrast,
$H_{\mathrm{sys}}''(t)$ for hybrid--Redfield contains off--diagonal elements.  Within
the Runge--Kutta scheme, this obstacle is easy to overcome, though it requires
diagonalization of the time dependent $H_{\mathrm{sys}}''(t)$ at every time step.
Because numerical diagonalization at every time--step is necessary for systems
with more than two degrees of freedom, this can become computationally
expensive for sufficiently large systems. 

\bibliography{library}    

\end{document}